\documentclass[11pt,UKenglish, autoref, thm-restate,a4paper]{amsart}
\pdfoutput=1
\usepackage[T1]{fontenc}
\usepackage{hyperref}
\usepackage{a4wide}
\usepackage{amsmath, amsthm, amssymb, graphicx, verbatim, color, esvect, complexity}
\usepackage[dvipsnames]{xcolor}
\usepackage{cleveref}
\definecolor{linkblue}{named}{MidnightBlue}
\hypersetup{colorlinks=true, linkcolor=linkblue, anchorcolor=linkblue,
  citecolor=linkblue, filecolor=linkblue, menucolor=linkblue,
  urlcolor=linkblue} 

\let\emph\relax 
\DeclareTextFontCommand{\emph}{\color{violet}\it}

\DeclareMathOperator{\conv}{conv}
\newtheorem{theorem}{Theorem}
\newtheorem{proposition}[theorem]{Proposition}
\newtheorem{lemma}{Lemma}[section]
\newtheorem{corollary}[theorem]{Corollary}

\newcommand{\RR}{\ensuremath{\mathbb{R}}}
\newcommand{\NN}{\ensuremath{\mathbb{N}}}

\newcommand{\g}{\gamma_0}

\newcommand{\vc}[1]{{#1}}
\newcommand{\dual}[1]{\widehat{#1}}
\newcommand{\pth}[1]{\left(#1\right)}
\newcommand{\ptb}[1]{\left\{#1\right\}}
\newcommand{\ext}{\mathrm{ext}} 
\newcommand{\Hit}[1]{\ensuremath{\mathcal{H}_P\pth{ {#1}}}} 
\makeatletter
\newcommand{\Smile}{\scalebox{0.6} {$\smile$}}%
\newcommand{\lift}[1]{
  \mathrel{\mathop{#1}\limits^{
    \vbox to -1.2\ex@{\kern-2\ex@
    \hbox{$\Smile$}\vss}}} \hspace{-3pt}}
\makeatother

\newcommand{\df}{\ell} 
\newcommand{\ds}{\delta} 

\title[Hitting and Covering Affine Families of Convex Polyhedra]{Hitting and Covering Affine Families of Convex Polyhedra, with Applications to Robust Optimization}

\author[J. Cardinal \and X. Goaoc \and S. Wajsbrot]{Jean Cardinal$^1$ \and Xavier Goaoc$^2$ \and Sarah Wajsbrot$^2$}
\address{$1$. Université libre de Bruxelles (ULB), Brussels, Belgium\\
  \texttt{jean.cardinal@ulb.be}}
\address{$2$. Université de Lorraine, CNRS, INRIA, LORIA, Nancy, F-54000, France\\
  \texttt{xavier.goaoc@loria.fr}, \texttt{sarah.wajsbrot@loria.fr}}

\begin{document}

\begin{abstract}
  Geometric hitting set problems, in which we seek a smallest set of points that collectively hit a given set of ranges, are ubiquitous in computational geometry. Most often, the set is discrete and is given explicitly. We propose new variants of these problems, dealing with continuous families of convex polyhedra, and show that they capture decision versions of the two-level finite adaptability problem in robust optimization. We show that these problems can be solved in strongly polynomial time when the size of the hitting/covering set and the dimension of the polyhedra and the parameter space are constant. We also show that the hitting set problem can be solved in strongly quadratic time for one-parameter families of convex polyhedra in constant dimension. This leads to new tractability results for finite adaptability that are the first ones with so-called left-hand-side uncertainty, where the underlying problem is non-linear.
\end{abstract}

\maketitle

\section{Introduction}

In this paper we present three contributions on hitting and covering problems for families of convex polyhedra. For the rest of the paper, the polyhedra we consider are all convex. First, we introduce a new flavour of these problems, dealing with hitting or covering {\em continuous} families of polyhedra, and show that they capture decision versions of the two-level finite adaptability problem in robust optimization. Next, we show that general methods from quantifier elimination yield strongly polynomial time algorithms for these problems when certain parameters are constant. Last, we present a fixed-parameter algorithm for a special case of our continuous hitting set problem.

\subsection{Hitting affine families of polyhedra}

\begin{figure}[!h]
\includegraphics[width=6.3cm,keepaspectratio]{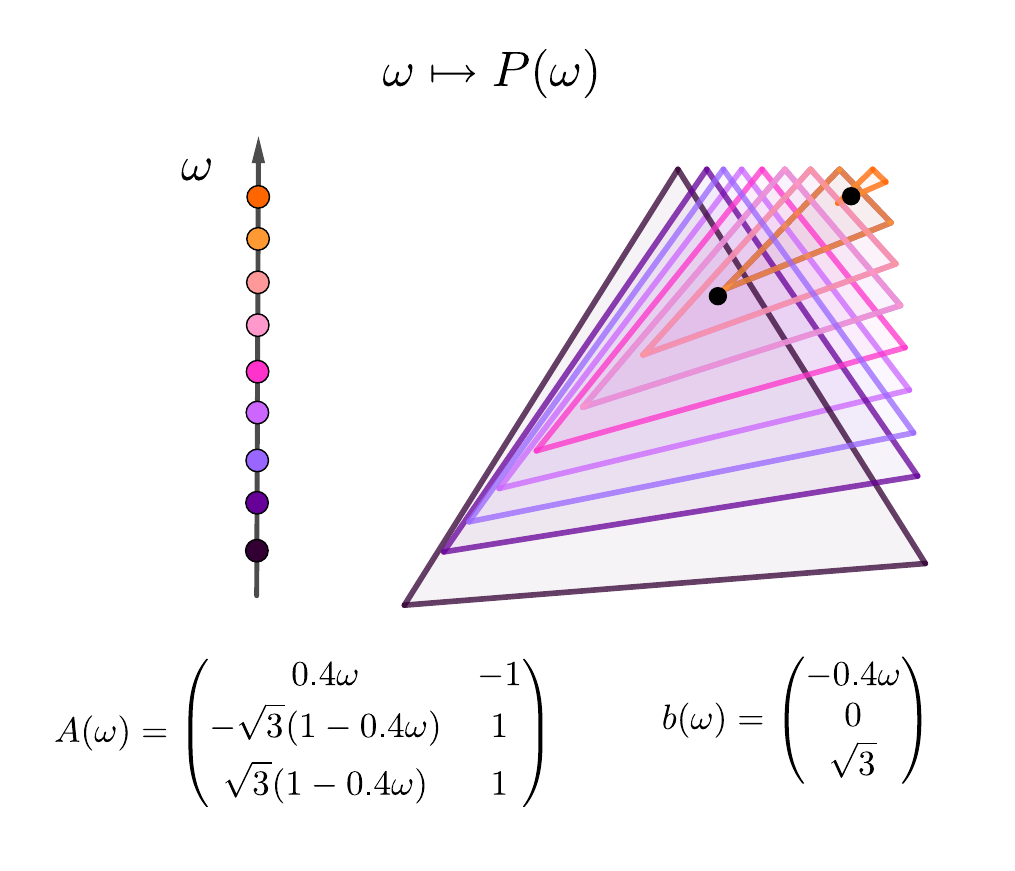}\hspace{7pt}\includegraphics[width=7.4cm,keepaspectratio]{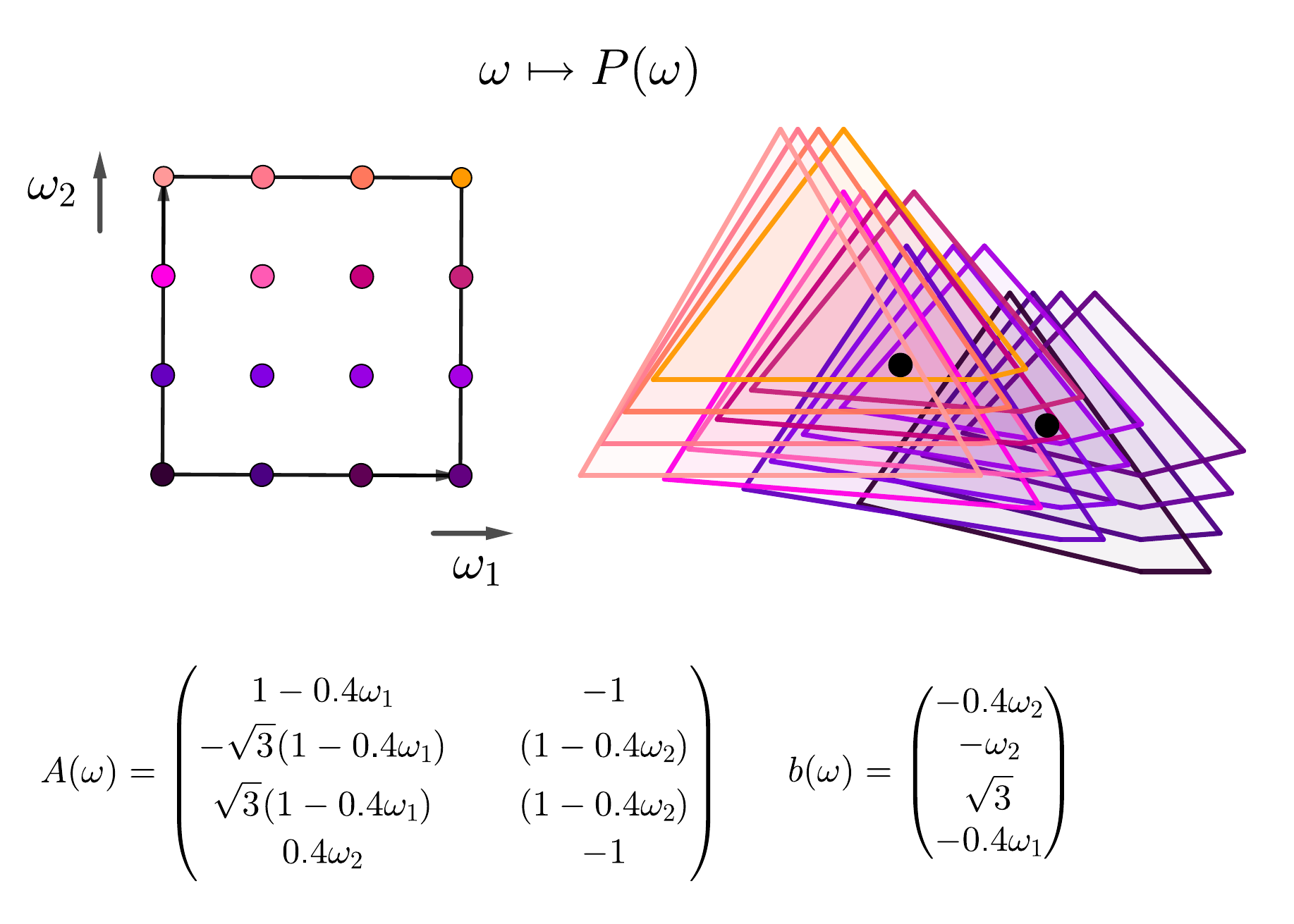} \qquad
  \caption{Two examples of affine families of polygons together with a hitting set (the black dots); For readability, we represent the polygon $P(\omega)$ for only a finite subset of $\omega$ in the domain. 
  \label{fig:example-hit}}
\end{figure}

An \emph{affinely parameterized family of polyhedra} in~$\RR^d$, or an
\emph{affine family of polyhedra} for short, is a continuous family
$P(\Omega)$ of polyhedra in $\RR^d$ defined by a domain $\Omega
\subset \RR^p$ and two affine maps $A:\Omega\mapsto
\mathbb{R}^{m\times d}$ and $\vc b:\Omega\mapsto \mathbb{R}^m$ via
\begin{equation}\label{eq:affpol}
  P(\Omega) = \left\{ P(\omega) \colon \omega \in \Omega\right\} \quad \text{where} \quad P(\omega) = \left\{\vc x\in \RR^d \colon A(\omega)\vc x \le \vc b(\omega) \right\}.
\end{equation}
Observe that an affine family of polyhedra has three defining
parameters: the dimension $d$ of the ambient space of the polyhedra,
the dimension $p$ of the parameter space $\Omega$, and the number $m$
of constraints defining each polyhedron. When every polyhedron
$P(\omega)$ is bounded, we call $P(\Omega)$ an \emph{affine family of
polytopes}. See Figure~\ref{fig:example-hit} for some examples.

We call a set $S \subset \RR^d$ a \emph{hitting set} for an affine
family of polyhedra if $S$ intersects every member of that family. See
Figure~\ref{fig:example-hit} for some examples.  We adopt the
following important convention: if an affine family of polyhedra
$P(\Omega)$ has one empty member, that is if $P(\omega) = \emptyset$
for some $\omega \in \Omega$, then $P(\Omega)$ has no hitting set. We
first consider the following computational problem:

\begin{quote}
  \textit{Given an integer $k$ and an affine family of polyhedra, does that family admit a hitting set of size $k$?}
\end{quote}

\subsection{Context: robust optimization and finite adaptability}

This continuous hitting set problem arises as the decision version of a special case of the  \emph{finite adaptability} problem in robust optimization~\cite{ben2009robust}. This line of research in mathematical programming deals with uncertainty in planning by modeling the decision to make as the optimization of some objective function under some constraints, with the objective function and the constraints depending not only of the free variables, but also of some uncertainty parameter~$\omega$. One then searches for an optimal solution that is valid for all values of the uncertainty parameter~$\omega$. Among robust optimization problems, those that involve succesive stages of decision are of particular interest~\cite{YANIKOGLU2019799}; Here is for instance the \emph{two-stage robust optimization problem}:
\begin{equation}\label{eq:ts}
  \begin{aligned}
    \inf_{\vc{x_f} \in \RR^{\df}} \quad \sup_{\omega \in \Omega} \quad & \inf_{\vc{x_s} \in \RR^{\ds}} \qquad {c_f}^T\vc{x_f} + c_s(\omega)^T \vc{x_s}\\
    & \text{s.~t.} \qquad A_f(\omega)\vc{x_f} + A_s(\omega)\vc{x_s}\le \vc b(\omega)
  \end{aligned}
\end{equation}
where $\Omega$ is the domain of uncertainty, $A_f(\omega)$, $A_s(\omega)$, $\vc b(\omega)$ and $\vc{d}(\omega)$ are input matrices and vectors depending on the uncertainty parameter $\omega$, and $\vc{c}$ is a deterministic input vector.
The variables $\vc{x_f}$ and $\vc{x_s}$ correspond respectively to the first and second stages of the optimization, where the second stage takes place only once the uncertainty $\omega$ has been revealed.
The problem is therefore to optimize in the first stage the worst-case outcome from the second stage.

\smallskip

In 2010, Bertsimas and Caramanis~\cite{bertsimas2010finite} proposed to approximate Problem~\eqref{eq:ts} by solving the following \emph{finite adaptability} problem:
\begin{equation}\label{eq:fa}
\begin{aligned}    
  \inf_{\substack{\vc{x_f} \in \RR^\df\\ \vc{x_s^{1}},\vc{x_s^{2}},\ldots, \vc{x_s^{k}} \in \RR^\ds}} \quad \sup_{\omega \in \Omega} \quad & \inf_{i \in [k]} \qquad {c_f}^T\vc{x_f} + c_s(\omega)^T \vc{x_s^i}\\
  &  \text{s.~t.} \qquad A_f(\omega)\vc{x_f} + A_s(\omega)\vc{x_s^{i}} \le \vc b(\omega)
\end{aligned}
\end{equation}
(We use $[k]$ to denote the set $\{1,2,\ldots, k\}$.) For fixed $k$ this problem is called the \emph{$k$-adaptability problem}. It models the precomputation, in the first stage, of $k$ candidate values for the second-stage variable $\vc{x_s}$. Once the uncertainty $\omega$ is revealed, the second stage consists of selecting one of the $k$ precomputed values that satisfies the constraints and minimizes the objective. Under a suitable continuity assumption, the value of Problem~\eqref{eq:fa} converges to the value of~\eqref{eq:ts} as $k\to\infty$~\cite[$\mathsection$~2]{kedad2023continuity}. See~\cite{HanasusantoKW15,PostekH16,BuchheimK17,subramanyam2020k} for recent work on Problem~\eqref{eq:fa} and variants.

In what follows, we will assume that $A_f$, $A_s$, $b$, and $c_s$ are affine maps, and consider only this version of the finite adaptability problem.

\subsection{Our results}

\noindent
{\em Problem correspondence.} The continuous hitting set problem stated above corresponds to the decision version of the optimization Problem~\eqref{eq:fa} when there is no first-stage decision ($\df = 0$); See Lemma~\ref{lem:decision-hit-nofd}. This correspondence is also the motivation for our convention that an affine family of polyhedra with one empty member has no hitting set, as such a family corresponds to an unfeasible problem. Like in the discrete setting, the hitting and covering problems for affine families of polyhedra enjoy some form of duality (see Section~\ref{s:hc-duality}). We prove (Lemma~\ref{lem:decision-hit}) that the decision version of the general finite adaptability problem is a special case of the following covering problem:

\begin{quote}
  \textit{Given two affine families of polyhedra in $\RR^d$ and  $k \in \NN$, does there exist $k$ polyhedra in the second family whose union covers a polyhedron from the first family?}
\end{quote}

\noindent
Notice that this problem is purely linear, whereas the finite adaptability Problem~\eqref{eq:fa} exhibits some non-linearity.

\medskip

\noindent
{\em Complexity.} We then turn our attention to the computational complexity of these problems. We first observe that quantifier elimination methods yield strongly polynomial solutions for these problems in fixed dimension.

\begin{theorem}
  \label{thm:appli}
  For every constant $k$, $\df$, $\ds$, $p$, there is an algorithm that solves the decision version of the $k$-adaptability Problem~\eqref{eq:fa} on $m$ constraints, with $\Omega$ a polyhedron in $\RR^p$ given as an intersection of $v$ halfspaces, in time strongly polynomial in $m$ and $v$.
\end{theorem}

\noindent
(see Apppendix~\ref{app:spa} for background on strongly polynomial algorithms.) Similarly, the hitting and covering problems stated above can be solved in time $m^{O(1)}$ when we fix $k$, the ambient dimensions of the polyhedra and the dimension of the parameter space(s) (see Section~\ref{sec:poly}). The bound in the exponent grows with the values of the fixed parameters, and the fixed-parameter tractability of these problems is a natural question. In the special case where the parameter domain is one-dimensional, we can give an algorithm for the hitting set problem where the order of the complexity does not depend on $k$ or the ambient dimension of the polyhedra.

\begin{theorem}\label{thm:onepara}
  For every constant $d$ and $k$, there is an algorithm that decides
  in strongly $O(m^2)$ time, given an affine family of polytopes
  $P([\alpha,\beta])$ in $\RR^d$, each defined by at most $m$
  constraints, whether $P([\alpha,\beta])$ admits a hitting set of
  size $k$.
\end{theorem}

\noindent
If $k$ is part of the input, then our algorithm has complexity $O(km^2)$ in the Real-RAM model (Proposition~\ref{prop:oneparaReal}) but is no longer strongly polynomial. 
We can extend the method behind Theorem~\ref{thm:onepara} to the optimization problem:

\begin{corollary}
  \label{cor:appli}
  For every constant $\ds$ and $k$, there is an algorithm that solves
  the $k$-adaptability Problem~\eqref{eq:fa}
  in the special case where $\ell=0$ and
  $\Omega$ is an interval, in time strongly $O(m^4)$.
\end{corollary}

\subsection{Background and related works}

\subsubsection{Complexity of finite adaptability}\label{s:previous}

Bertsimas and Caramanis
proved~\cite[Prop. 5]{bertsimas2010finite} that unless $\textsc{P} = \textsc{NP}$,
there is no polynomial-time algorithm for the finite adaptability
problem~\eqref{eq:fa} already in the special case where $k=2$, there
is no first decision $x_f$, and $A_s$ is constant, \textit{i.e.}
independent of $\omega$.  Their proof requires, however, that the
dimensions of $\vc{x_s}$ and $\omega$ as well as the number $m$ of
rows in $A_s$ be part of the input. A natural question is whether the
computational complexity becomes polynomial when some of these
parameters are constant.

We are only aware of two\footnote{Note that Bertsimas and Caramanis gave a reformulation of Problem~\eqref{eq:fa} as a bilinear optimization problem with a linear objective function~\cite[Prop. 4]{bertsimas2010finite}, but it does not yield a polynomial-time algorithm even when all dimensions and $k$ are fixed. Indeed, their reformulation uses $\ell+d+m^k$ variables and $m^k(v+2)$ bilinear constraints, where $v$ is the number of extreme points of the domain $\Omega$.} previous tractability results for Problem~\eqref{eq:fa}.
On the one hand, Subramanyam, Gounaris and Wiesemann~\cite[Prop. B.3]{subramanyam2020k} observed that Problem~\eqref{eq:fa} reduces to linear programming, and is therefore polynomial, when $A_f$, $A_s$ and $b$ are constant (but $c_s$ still depends on $\omega$).
On the other hand, Kedad-Sidhoum, Meunier and Medvedev~\cite[Theorems~1.1 and~1.2]{kedad2023continuity} proved that if $k \le 3$, the number of vertices (and dimension) of $\Omega$ is bounded, and only the right-hand side $\vc b$ of the constraints depends on the uncertainty parameter (that is, $A_f$,
$A_s$ and~$\vc d$ are constant), then Problem~\eqref{eq:fa} reduces to a constant number of linear programs and is therefore polynomial in the Turing machine model.

Theorem~\ref{thm:appli} therefore identifies new tractable cases of the decision version of the finite adaptability Problem~\eqref{eq:fa}, and Corollary~\ref{cor:appli} shows that without first decision and with one-dimensional uncertainty, the problem can be solved in $O(m^4)$ time for any constant $k$ and $\ds$. Let us stress that these are the first tractability results with so-called {\em left-hand-side uncertainty}, that is with the map $A_s$ affine rather than constant, which makes the underlying problem non-linear in $\vc{x_f}, \vc{x_s}$ and~$\omega$.

\subsubsection{Related notions in computational geometry.}

Hitting set problems were extensively investigated in discrete and computational geometry for discrete, unstructured families of sets.
One typically expects that deciding whether $k$ points suffice to hit $n$ given subsets of $\RR^d$ is \textsc{NP}-hard when $k$ or $d$ is part of the input;
see for instance the landmark results of Megiddo~\cite{megiddo1990complexity} and Megiddo and Supowit~\cite{MS84}.
When both $k$ and $d$ are fixed, the problem can be solved in polynomial time via the computation of an arrangement.

Notions of uncertainty related to robust mathematical programming were also investigated by computational geometers.
In this context, uncertainty in the input geometric data is typically modeled by replacing single points by regions (say, disks of some small radius), in which the actual point is guaranteed to lie.
One can preprocess the regions, and prepare for the worst-case actual input.
Recent examples include sorting and finding the Pareto front of imprecise point sets~\cite{HKLS22,HKLS24}.
We refer to L\"{o}ffler for a general discussion of \emph{imprecision in computational geometry}~\cite{L09}.

More generally, continuous families of geometric sets are ubiquitous in computational geometry.
In \emph{range searching}, for instance, we are given a collection of points in $d$-dimensional space, and are asked to construct a data structure that allows to answer queries of the following form:
Given a \emph{range} $R\subset \RR^d$, report the set of points from the collection that are contained in $R$.
The ranges are typically restricted to be axis-aligned boxes, halfspaces, or semialgebraic sets, giving rise to the corresponding \emph{orthogonal}, \emph{halfspace}, or \emph{semialgebraic} range searching problems; see the survey of Agarwal~\cite{agarwal2017range} for references.
In the problem of finding small \emph{$\varepsilon$-net}, the goal is similar to ours, in that we wish to find small hitting sets of continuous family of ranges.
More precisely, given a finite set $S$ of points together with a continuous family of ranges (typical examples are all halfspaces, or all unit disks), we wish to find a set $N$ of points that collectively hit all ranges that contain at least an $\varepsilon$ fraction of the points of $S$~\cite{HW87}.
If the set is restricted to be a subset of $S$, then $N$ is called a \emph{strong $\varepsilon$-net}, and a \emph{weak $\varepsilon$-net} otherwise~\cite{CEGGSW95}.
While this also bears a superficial resemblance with our question, it is yet quite different, since in our case, all members of the input continuous family have to be hit, whatever their size is.
Br\"{o}nimann and Goodrich~\cite{BG95}, and then Even, Rawitz, and Shahar~\cite{ERS05}, showed that finding $\varepsilon$-nets of size $O(c/\varepsilon )$ in polynomial time implied the existence of $c$-approximation algorithms for the discrete hitting set problem.
In the same spirit, Elbassioni~\cite{elbassioni2023bicriteria} recently proposed an approximation algorithm for finding small hitting sets of infinite range spaces of bounded VC-dimension, and gave an application to the problem of covering a polygonal region by translates of a given polygon.


\section{Relation between hitting/covering and finite adaptability}

Let us clarify the relation between the hitting and covering problems for affine families of polytopes and the finite adaptability Problem~\eqref{eq:fa}. We assume in this section that the domain of uncertainty $\Omega$ is a polyhedron in $\RR^p$, the matrices $A_f(\omega)$, $A_s(\omega)$ and the vectors $\vc b(\omega)$, $c_s(\omega)$ depend affinely on the uncertainty parameter $\omega$, and the vector $c_f$ is deterministic.

\subsection{The hitting-covering duality}
\label{s:hc-duality}

Let us start by remarking that for any regions $\Lambda \subseteq \Omega \subseteq \RR^p$, every affine map on $\Omega$ restricts, in a unique way, to an affine map on $\Lambda$; Conversely, every affine map on $\Lambda$ extends to an affine map on $\Omega$, and this extension is unique if $\Omega$ is contained in the affine span of $\Lambda$. This implies that any affine family of polyhedra $P(\Omega)$ restricts (in a unique way) to an affine family of polyhedra $P(\Lambda)$, and that conversely every affine family of polyhedra $P(\Lambda)$ extends to an affine family of polyhedra $P(\Omega)$, and that this extension is unique if $\Omega$ is contained in the affine span of $\Lambda$.

Let us fix two affine maps $A:\RR^p\mapsto \mathbb{R}^{m\times d}$ and $b:\Omega\mapsto \mathbb{R}^m$ and consider the maps
\[\begin{aligned}
P\colon & \omega \mapsto \ptb{x \in \RR^d \colon A(\omega) x \le b(\omega)} & \text{ for } \omega \in \RR^p,\\
  \dual P\colon & x \mapsto \ptb{\omega \in \RR^p \colon A(\omega) x \le b(\omega)} & \text{ for } x \in \RR^d,
\end{aligned}\]
so that $x \in P(\omega)$ if and only if $\omega \in \dual P(x)$.
Hence, switching from $P$ to $\dual P$ exchanges hitting and covering in the sense that for any domain $\Omega \subseteq \RR^p$, a set $S \subseteq \RR^d$ is a hitting set for $\{P(\omega) \colon \omega \in \Omega\}$ if and only if $\Omega \subseteq \cup_{x \in S} \dual P(x)$.
It turns out that $\dual P$ is also an affine family of polyhedra, which we call the \emph{dual} of $P$.

\begin{lemma}\label{l:dual}
  If $P(\RR^p)$ is an affine family of polyhedra in $\RR^d$, each defined by $m$ constraints, then $\dual P(\RR^d)$ is an affine family of polyhedra in~$\RR^p$, each defined by $m$ constraints.
\end{lemma}
\begin{proof}
  Let us fix a basis $(e_1,e_2, \ldots, e_p)$ of $\RR^p$, so as to write $\omega = (\omega_1,\omega_2, \ldots, \omega_p)$, $A(\omega)= A_0 + \sum_{i=1}^p \omega_i A_i$ and $b(\omega)= b_0+ \sum_{i=1}^p \omega_i b_i$, where $A_0, A_1, \ldots, A_p\in  \RR^{m\times d}$, $\vc b_0, \vc b_1, \ldots, \vc b_p\in \RR^m$. For $x \in \RR^d$, we have
  \[\begin{aligned}
  \dual P(x) & = \left\{\omega \in \RR^p \colon \pth{A_0+\sum_{i=1}^p \omega_i A_i} x \le b_0 + \sum_{i=1}^p \omega_i b_i\right\}\\
  & = \left\{\omega \in \RR^p \colon \sum_{i=1}^p  \pth{A_i x - b_i} \omega_i \le b_0 -A_0\vc{x}\right\}.
  \end{aligned}\]
  Let $\dual A (\vc x) = \sum_{i=1}^p \pth{A_i x - b_i} {e_i}^T \in \RR^{m \times p}$, so that  $\dual A (x)\omega =  \sum_{i=1}^p \pth{A_i x - b_i} \omega_i$. Let $\dual b (x)=  b_0 -A_0 x$. Note that $\dual P(x) = \{\omega \in \RR^p \colon \dual A (\vc x) \omega \le \dual {\vc b} (\vc x)\}$ with $\dual A$ and $\dual {\vc b}$ affine maps. It follows that  $\dual P (\RR^d)$ is an affine family of polyhedra, with ambient space of dimension $p$, parameter space of dimension $d$ and each polyhedron defined by $m$ constraints.
\end{proof}

\noindent
When $\Omega$ is a polyhedron in $\RR^p$, a natural candidate for the dual of an affine family $P(\Omega)$ of polyhedra is $\{\omega \in \Omega \colon x \in P(\omega)\}$.
This is again an affine family of polyhedra, which we denote by $\dual P_{|\Omega}(\RR^d) = \{\dual P(x) \cap \Omega \colon x \in \RR^d\}$.
In other words, restricting $P(\RR^p)$ into $P(\Omega)$ translates, under duality, into intersecting each dual polyhedron with $\Omega$.

\subsection{Direct correspondences}\label{s:twostageform}

Let us first consider the $k$-adaptability problem~\eqref{eq:fa} without first decision ($\df = 0$), that is
\begin{equation}\label{eq:fanofd}
\begin{aligned}    
  \inf_{\vc{x_s^{1}},\vc{x_s^{2}},\ldots, \vc{x_s^{k}} \in \RR^{\ds}} \quad \sup_{\omega \in \Omega} \quad & \inf_{i \in [k]} \qquad c_s(\omega)^T \vc{x_s^i}\\
  &  \text{s.~t.} \qquad A_s(\omega)\vc{x_s^{i}} \le \vc b(\omega)
\end{aligned}
\end{equation}
\begin{lemma}\label{lem:decision-hit-nofd}
  For any real $t$, the value of the $k$-adaptability problem without first decision~\eqref{eq:fanofd} is at most  $t$ if and only if the affine family of polyhedra $P_t(\Omega) = \{P_t(\omega) \colon \omega \in \Omega\}$ defined by
\[ P_t(\omega) = \left\{\vc x\in \RR^d \colon \pth{\begin{matrix} A_s(\omega)\\ c_s(\omega)^T
\end{matrix}} \vc x \le \pth{\begin{matrix} \vc b(\omega)\\t\end{matrix}} \right\}\]
admits a hitting set of size $k$.
\end{lemma}
\begin{proof}
  Let $t\in \RR$.
  The value of Problem~\eqref{eq:fanofd} is at most $t$ if and only if there exists $\vc{x_s} = (\vc{x_s^{1}},\vc{x_s^{2}},\ldots, \vc{x_s^{k}}) \in (\RR^d)^k$ such that $\sup_{\omega \in \Omega} \inf_{i \in [k] \text{ s.t. } A_s(\omega)\vc{x_s^{i}}\le \vc b(\omega)} c_s(\omega)^T \vc{x_s^i}(\omega)$ is at most~$t$. This is equivalent to the condition that for every $\omega \in \Omega$, $\inf_{i \in [k] \text{ s.t. } A_s(\omega)\vc{x_s^{i}}\le \vc b(\omega)} c_s(\omega)^T \vc{x_s^i}(\omega)$ is at most~$t$. This, in turn, is equivalent to the condition that for every $\omega \in \Omega$, there exists $i \in [k]$ such that $A_s(\omega) \vc{x_s^{i}} \le \vc b(\omega)$ and $c_s(\omega)^T \vc{x_s^i}(\omega) \le t$;
  In other words, $\vc{x_s^i} \in P_t(\omega)$.
  Hence, the value of Problem~\eqref{eq:fanofd} is at most $t$ if and only if $P_t(\Omega)$ admits a hitting set of size $k$.
  Observe that in particular, Problem~\eqref{eq:fanofd} is infeasible if and only if there is no feasible solution $(\vc{x_s^{1}},\vc{x_s^{2}},\ldots, \vc{x_s^{k}})$, that is no hitting set of size~$k$ for $P(\Omega)$.
\end{proof}

When there is a first decision ($\df>0$), one may proceed as in Lemma~\ref{lem:decision-hit-nofd}, \textit{mutatis mutandis}, and obtain the following characterizations of the fact that the value of~\eqref{eq:fa} is at most $t$:
\begin{itemize}
\item[(a)] The affine family of polyhedra $P_t(\Omega) = \{P_t(\omega) \colon \omega \in \Omega\}$ defined by
\[ P_t(\omega) = \left\{\pth{\begin{matrix}\vc{x_f}\\\vc{x_s}
  \end{matrix}} \in \RR^{\df+\ds} \colon \pth{\begin{matrix} A_f(\omega) & A_s(\omega)\\ {c_f}^T & c_s(\omega)^T
\end{matrix}} \pth{\begin{matrix}\vc{x_f}\\\vc{x_s} \end{matrix}} \le \pth{\begin{matrix} \vc b(\omega)\\t\end{matrix}} \right\}\]
admits a hitting set of size $k$ in which all points have the same first $\df$ coordinates.

\item[(b)] There exists $\vc{x_f} \in \RR^\df$ such that the affine family of polyhedra $P_{\vc{x_f},t}(\Omega)= \{P_{\vc{x_f},t}(\omega) \colon \omega \in \Omega\}$ defined by
\[ P_{\vc{x_f},t}(\omega) = \left\{\vc{x_s} \in \RR^{\ds} \colon \pth{\begin{matrix} A_s(\omega)\\ c_s(\omega)^T
\end{matrix}}\vc{x_s} \le \pth{\begin{matrix} \vc b(\omega)-A_f(\omega)\vc{x_f}\\t-{c_f}^T\vc{x_f}\end{matrix}} \right\}\]
admits a hitting set of size $k$.
\end{itemize}
In other words, we have to either require the hitting set to lie on some coordinate subspace (a), or work with an affinely parameterized family of affine families of polyhedra (b). Each affine family of polyhedra in~(b) corresponds to the slice of the affine family of polyhedra of~(a) by a coordinate subspace of  codimension $\df$.

\subsection{A simpler correspondence via lifting}\label{s:lift}

We now establish that comparing the value of the $k$-adaptability problem~\eqref{eq:fa} to a given value $t$ is equivalent to a problem of covering one member of an affine family of polyhedra by $k$ members of another family.
To obtain this reformulation, we fix $\ell$ and $p$ and, taking inspiration from the Veronese map, define a ``lifting'':
\[L: \left\{\begin{array}{rcl} \RR^{\df+p} & \to & \RR^{\df+p+\df p}\\(\vc{x_f}, \underbrace{\omega_1, \omega_2, \ldots, \omega_p}_{\omega})&\mapsto&(\vc{x_f}, \omega, \omega_1 \vc{x_f},\omega_2 \vc{x_f}, \ldots, \omega_p \vc{x_f}).\end{array}\right.\]
For better readability, we let $d = \df+p+\df p$. We use $\vc z$ to denote a point in $\RR^d$, and write $z_{x_f} = (z_1,\ldots,z_{\df})^T $ and $z_{\omega} = (z_{\df+1},\ldots,z_{\df + p})^T $. For $\omega \in \Omega$, we decompose $A_f(\omega)$ into $A_f(\omega)= A_{L,0} + \sum_{i\in[p]} \omega_i A_{L,i}$, where $A_{L,0},A_{L,1}, \ldots, A_{L,p} \in \RR^{m\times \ds}$. 

\begin{lemma}\label{lem:decision-hit}
  For any real $t$, the value of the $k$-adaptability problem~\eqref{eq:fa} is at most $t$ if and only if there is a member of $\lift {P_t}(\RR^\df)$ that can be covered by some $k$ members of $\lift {Q_t}(\RR^\ds)$, where $\lift {P_t}(\RR^\df)$ is the affine family of polyhedra in $\RR^d$ defined by $\lift{P_t}(\RR^\df) = \{\lift {P_t}(x_f) \colon x_f \in \RR^\df\}$ where $\lift {P_t}(x_f) = L(x_f \times \Omega)$, and $\lift {Q_t}(\RR^\ds)$ is the affine family of polyhedra in $\RR^d$ defined by $Q_t(\RR^\ds) = \{\lift {Q_t}(x_s) \colon x_s \in \RR^\ds\}$ where
  \[ \lift {Q_t}(x_s) = \left\{ z \in \RR^d \colon \left\{
  \begin{aligned}
    A_{L,0} z_{x_f} + A_s(z_\omega) x_s + \sum_{i\in [p]} A_{L,i} (z_{ip+\df +1},\ldots,z_{ip+\df +p})^T &\le b(z_\omega)\\ \vc
{c_f}^T \vc z_{x_f} + c_s(\vc z_\omega)\vc x_s &\le t
\end{aligned} \right.
 \right\}.\]
\end{lemma}
\begin{proof}
  Let us first note that for any fixed $t \in \RR$, $\lift {P_t}(\RR^\df)$ and $\lift {Q_t}(\RR^\ds)$ are indeed affine families of polyhedra. For $Q_t(\RR^\ds)$, this follows from the assumption that $A_s(\omega)$ depend affinely on $\omega$ and $A_{L,0},A_{L,1}, \ldots, A_{L,p}$ are constant matrices. For $\lift {P_t}(\RR^\df)$, this follows from the assumption that $\Omega$ is a polyhedron.

  Next, observe that $L(\RR^{\df+p})$ is the surface $\Sigma \subseteq \RR^d$ of dimension $\df+p$ defined by the quadratic equations $z_{p+i\df+j} = z_{j}z_{\df+i}$ for $1 \le i \le p$ and $1 \le j \le \df$. Letting $\pi:(z_1,z_2, \ldots, z_{d}) \mapsto (z_1,z_2, \ldots, z_{\df+p})$ be the projection that forgets the last $\df p$ coordinates, we see that $L$ is a bijection with inverse $\pi_{|\Sigma}$.

  Let us fix $t\in \RR$ and start from the characterization (b) from Section~\ref{s:twostageform} -- from which we recover the notation $P_{x_f,t}(\omega)$. The value of the $k$-adaptability problem~\eqref{eq:fa} is at most $t$ if and only if there exists $(x_f,x_s^1, x_s^2, \ldots,\vc x_s^k) \in \RR^\df \times (\RR^\ds)^k$ such that for every $\omega \in \Omega$, there exists $i \in [k]$ such that $x_s^i \in P_{x_f,t}(\omega)$. Letting $S_t(x_s)= \ptb{(x_f,\omega) \in \RR^\df \times \Omega \colon x_s \in P_{x_f,t}(\omega)}$, the characterization becomes: there exists $(x_f,x_s^1, x_s^2, \ldots,\vc x_s^k) \in \RR^\df \times (\RR^\ds)^k$ such that $\{x_f\} \times \Omega$ is covered by $S_t(x_s^1) \cup S_t(x_s^2) \cup \ldots \cup S_t(x_s^k)$.

  Let us now consider how the lift $L$ acts on this characterization. First, $L$ is a bijection from $\RR^{\df+p}$ to $\Sigma$, so $\{x_f\} \times \Omega$ is covered by  $S_t(x_s^1) \cup S_t(x_s^2) \cup \ldots \cup S_t(x_s^k)$ if and only if $L(\{x_f\} \times \Omega)$ is covered by $L(S_t(x_s^1)) \cup L(S_t(x_s^2)) \cup \ldots \cup L(S_t(x_s^k))$. Now,  $L(\{x_f\} \times \Omega) = \lift P(x_f)$ by definition of $\lift P $. On the other hand, the definition of $\lift Q $ ensures that $L(S_t(x_s))=\lift {Q_t}(x_s) \cap \Sigma$. The characterization therefore reformulates as: there exists $(x_f,x_s^1, x_s^2, \ldots,\vc x_s^k) \in \RR^\df \times (\RR^\ds)^k$ such that $\lift P(x_f)$ is covered by  $\pth{\lift {Q_t}(x_s^1) \cap \Sigma} \cup \pth{\lift {Q_t}(x_s^2) \cap \Sigma} \cup \ldots \cup \pth{\lift {Q_t}(x_s^k) \cap \Sigma}$. Since for every $x_f \in \RR^\df$, the set $\lift P(x_f)$ is contained in $\Sigma$, we can drop the intersections with $\Sigma$ in that characterization, and the statement follows.
\end{proof}


\section{Polynomial complexity bounds}\label{sec:poly}

We first show that algorithms for quantifier elimination yield, when the parameter $k$ and the dimensions are constant, strongly polynomial algorithms for the decision version of the $k$-adaptability Problem~\eqref{eq:fa} as well as for our continuous hitting and covering problems. 

\subsection{Semi-algebraic sets and first-order formulas}

Our proofs involve the analysis of families of polynomial inequalities in $\RR^d$, that is of semi-algebraic sets. It will be convenient to formulate these sets in the equivalent form of first-order formulas (see the book of Basu, Pollack and Roy~\cite[$\mathsection 2.3$]{basu:hal-01083587}). We will use the following complexity bounds on quantifier elimination.

\begin{theorem}[{\cite[$\mathsection 14$, Theorem 14.16]{basu:hal-01083587}}]\label{thm:elim}
  Let $\phi(X_1,X_2,\ldots,X_d)$ be a first-order formula of the form $(Q_1 \mathbf{Y}_1)(Q_2 \mathbf{Y}_2)\ldots(Q_n \mathbf{Y}_n)F(X,Y)$, where $Q_1$, $Q_2$, \ldots, $Q_n$ are quantifiers in $\{\forall,\exists\}$ and $F$ is a quantifier free formula. Assume that each set $\mathbf{Y}_{i}$ has at most $b$ variables and that $F$ is built with at most $s$ polynomials of degree at most $r$ (in $nb+d$ variables). 

  \begin{itemize}
  \item[(i)] There exists a quantifier free formula equivalent to
    $\phi(X_1,X_2, \ldots, X_d)$ involving at most~$\nu$ polynomials,
    each of degree at most $\delta$,

  \item[(ii)] this quantifier free formula can be computed in $\nu$
    arithmetic operations in the ring generated by the coefficients of
    the polynomials, and

  \item[(iii)] if the coefficients of the input polynomials are in
    $\mathbb{Z}$ and have bitsize at most $\tau$, then the polynomials
    that arise in the algorithm have coefficients of bitsize bounded
    by $\tau \delta$,
  \end{itemize}
  where $\nu = s^{(d+1)(b+1)^n}r^{dO(b)^n}$ and $\delta = r^{O(b)^n}$.
\end{theorem}

\noindent
We will also use the following complexity bound on the decision problem for the existential theory of the reals.

\begin{theorem}[{\cite[$\mathsection 13$, Theorem 13.13]{basu:hal-01083587}}]\label{thm:etr}
  Let $\phi$ be a quantifier free formula with $d$ free variables, built on $\nu$ polynomials of degree at most $\delta$.

  \begin{itemize}
  \item[(i)] The truth of the sentence $(\exists X \in \RR^d,\phi(X))$ can be decided in $\nu^{d+1}\delta^{O(d)}$ arithmetic operations in the ring generated by the coefficients of the polynomials, and 
  \item[(ii)] if the coefficients of the input polynomials are in $\mathbb{Z}$ and their bitsize is bounded by $\tau$, the polynomials that arise in the algorithm have coefficients of bitsize bounded by $\tau \delta^{O(d)}$.
  \end{itemize}
\end{theorem}

\subsection{Application to $k$-adaptability}

The decision version of Problem~\eqref{eq:fa} consists in deciding, given some real number $t$, whether the optimal value of \eqref{eq:fa} is at most $t$.

\begin{proof}[Proof of Theorem~\ref{thm:appli}]
  The decision version of Problem~\ref{eq:fa} can be cast as deciding the validity of the formula
  \[
  \exists x_f \in \RR^{\df}, x_s^1, x_s^2,\ldots ,x_s^k\in \RR^\ds\ \ \Phi(x_f,x_s),
  \]
  where
  \[
    \Phi(x_f, x_s) \equiv \forall\omega\in\RR^p\quad
    \pth{\omega\not\in\Omega} \vee 
    \pth{(x_f,x_s^1)\in P_t(\omega)}\vee 
    \pth{(x_f,x_s^2)\in P_t(\omega)}\vee 
    \ldots \vee \pth{(x_f,x_s^k)\in P_t(\omega)},
  \]
  and $P_t(\Omega)$ is the affine family of polyhedra defined by
  \[
  P_t(\omega) = \left\{
  (x_f, x_s) \in \RR^{\df+\ds} \colon
  \pth{\begin{matrix}
      A_f(\omega) A_s(\omega)\\
      c_f(\omega)^T c_s(\omega)^T
  \end{matrix}}
  (x_f, x_s) \le
  \pth{\begin{matrix}
      b(\omega)\\
      t
  \end{matrix}}
  \right\}.
  \]
  Note that the polyhedron $P_t(\omega)$ encode both the constraints enforced by $A_f,A_s$, and $b$ on $(x_f,x_s^i)$, and the bound $t$ on the optimal value (this is similar to the proof of Lemma~\ref{eq:fanofd}).
  The universal quantifier on $\omega$ takes care of the $\sup_{\omega \in \Omega}$ in the formulation~\eqref{eq:fa}, while the disjunction on the terms of the form $(x_f,x_s^i)\in P_t(\omega)$ takes care of the $\inf_{i \in [k]}$.
  
  We can apply Theorem~\ref{thm:elim} to eliminate the universal quantifier in $\Phi(x_f, x_s)$.
  Referring to the parameters in Theorem~\ref{thm:elim}, we have $n=1$ and $b=p$, and there are $s\le v+k(m+1)$ polynomials involved, defining the polyhedra $\Omega$ and $P_t(\omega)$.
  Each polynomial is of degree at most $r=2$, and the total number of variables is $k\ds +\df$.
  Hence, the elimination of the quantifier takes time $O\pth{\pth{v+m}^{(k\ds +\df+1)(p+1)}}$.
  
  We now have a formula with a single existential quantifier on $(x_f,x_s)$, to which we can apply Theorem~\ref{thm:etr}.
  Following Theorem~\ref{thm:elim}, the degree is at most $r^{O(b)^n} = 2^{O(p)}$, and we can therefore decide the validity of the formula in time
  \[
  \pth{O\pth{\pth{v+m}^{(k\ds +\df+1)(p+1)}}}^{k\ds +\df+1}2^{O((k\ds +\df)p)} = O\pth{\pth{v+m}^{(k\ds +\df+1)^2(p+1)}}
  \]
  as claimed.
\end{proof}

Similarly, we can express the hitting and covering problems for affine families of polyhedra as the decision of the validity of a formula, which can then be handled by Theorems~\ref{thm:elim} and~\ref{thm:etr}. This gives the following results:
\begin{itemize}
\item For every constant $k$, $p$, and $d$ there is an algorithm that decides in time strongly polynomial in $m$, given a polyhedron
  $\Omega$ in $\RR^p$ defined by $m$ constraints and an affine family of polyhedra $P(\Omega)$ in $\RR^d$, each defined by at most $m$
  constraints, whether $P(\Omega)$ admits a hitting set of size
  $k$. The complexity is $m^{O(1)}$, where the exponent depends on $k$, $p$, and~$d$.

\item For every constant $k$, $\gamma$, $\lambda$ and $d$ there is an algorithm that decides in time strongly polynomial in $m$, given an
  affine family of polyhedra $P(\Gamma)$ in $\RR^d$ defined by at most
  $m$ constraints, and an affine family of polyhedra $Q(\Lambda)$ in
  $\RR^d$, defined by at most $m$ constraints, with $\Gamma \subseteq
  \RR^\gamma$ and $\Lambda \subseteq \RR^\lambda$ polyhedra defined by
  at most $m$ constraints each, whether there exist $\g \in \Gamma$
  and $\lambda_1,\lambda_2, \ldots, \lambda_k \in \Lambda$ such that
  $P(\g ) \subseteq Q(\lambda_1) \cup Q(\lambda_2) \cup \ldots \cup
  Q(\lambda_k)$. The complexity is $m^{O(1)}$, where the exponent depends on $k$, $\gamma$, $\lambda$ and~$d$.
\end{itemize}

\noindent
We also note that similar tools~\cite[$\mathsection 14$, Algorithm 14.9]{basu:hal-01083587} can solve the optimization version of the $k$-adaptability Problem~\eqref{eq:fa}, with $\Omega$ a polyhedron in $\RR^p$, in polynomial time for every constant $k$, $\df$, $\ds$ and $p$.


\section{Hitting a one-parameter family of polyhedra}

We now prove Theorem~\ref{thm:onepara}: for every constant $d$,
the size $k$ of the smallest hitting set of a one-parameter affine
family of polytopes in $\RR^d$ can be computed in $O(km^2)$, where $m$
denotes the number of constraints defining each polytope. The proof is
essentially a greedy algorithm building on an application of
parametric search~\cite{megiddo1983applying} to a carefully chosen
linear-time algorithm for linear programming in fixed dimension, here
we use Megiddo's~\cite{megiddo1984linear}.

\subsection{The intersection of an affine family of polyhedra}\label{sec:hit1point}

We define the \emph{common intersection region} of an affine family of polyhedra $P(\Omega)$ in $\RR^d$ as the set $\Hit{\Omega} = \cap_{\omega \in \Omega} P(\omega)$.
We let $\ext(C)$ denote the set of extreme\footnote{Recall that a point $c$ in a convex $C$ is called \emph{extreme} if it cannot be expressed as the midpoint of two distinct points in $C$, or equivalently if $C\setminus\{c\}$ is convex~\cite[Chapter A, $\mathsection~2.3$, Definition 2.3.1]{hiriart-lemarechal-2001}.} points of a set $C\subset \RR^p$, and write $\overline X$ for the topological closure of a subset $X \subseteq \RR^p$.

\begin{lemma}\label{lem:extreme}
  For any affine family of polyhedra $P(\Omega)$ and for any subset
  $\Lambda \subseteq \Omega$ of its parameter set, we have
  (i) $\Hit{\Lambda} = \Hit{\conv \Lambda }$,
  (ii) $\Hit{\Lambda}  = \Hit{\overline{\Lambda}}$ and,
  (iii) if $\Lambda$ is bounded, $\Hit{\Lambda} = \Hit{\ext\pth{\overline{\conv \Lambda}}}$.
\end{lemma}

\begin{proof}
  Observe first that for every subsets $A,B \subseteq \Omega$, if $A\subseteq B$, then  $\bigcap_{\omega\in B} P(\omega) \subseteq \bigcap_{\omega\in A} P(\omega)$. It follows that $\Hit{\Lambda} \supseteq \Hit{\conv \Lambda}$ and $\Hit{\Lambda} \supseteq \Hit{\overline{\Lambda}}$.


Let us prove that $\Hit{\Lambda} \subseteq \Hit{\conv \Lambda}$. Let $\vc x \in \bigcap_{\omega\in \Lambda} P(\omega)$. Consider a finite convex combination $\omega = \sum_{i=1}^{v} \alpha_i \lambda_i$ where $v\in \mathbb{N}$ and $\lambda_1,\ldots,\lambda_v \in \Lambda$. By assumption, for all $i\in [v]$, $\vc x$ hits the polyhedron $P(\lambda_i)$, that is for all $i\in [v]$, $A(\lambda_i)\vc x \le b(\lambda_i)$. It follows that
\[ A(\omega)\vc x = \sum_{i=1}^v \alpha_i A(\lambda_i) \vc x \le \sum_{i=1}^v \alpha_i b(\lambda_i) = b(\omega),\]
and $\vc x \in P(\omega)$. This completes the proof of~(i).


 Let us now prove that $\Hit{\Lambda} \subseteq \Hit{\overline{\Lambda}}$. Let $\vc x \in \bigcap_{\omega\in \Lambda} P(\omega)$. Let $\omega \in \overline{\Lambda}$. Then there exists a sequence $(\lambda_n)_{n\in\mathbb{N}}$ of elements of $\Lambda$ that converges to $\omega$. By assumption, for all $n\in \mathbb{N}$, $\vc x$ hits the polyhedron $P(\lambda_n)$, that is for all $n\in \mathbb{N}$, $A(\lambda_n)\vc x \le b(\lambda_n)$. Since $A$ and $b$ are continuous, it comes that $A(\omega)\vc x \le b(\omega)$, that is $\vc x \in P(\omega)$. This completes the proof of~(ii).


 Let us prove $\Hit{\Lambda} = \Hit{\ext\pth{\overline{\conv \Lambda }}}$. A theorem of Minkowski (see~\cite[Chapter A, $\mathsection~2.3$, Theorem 2.3.4]{hiriart-lemarechal-2001}) asserts that if $C$ is compact, convex in a finite dimensional euclidean space, then $C$ is the convex hull of its extreme points. Hence, if $\Lambda$ is bounded, $\overline{\conv \Lambda}$ is a compact convex set in $\RR^p$, and is therefore the convex hull of its extreme points. So we have:
 \[\begin{aligned}
 \bigcap_{\omega\in \Lambda} P(\omega)  \stackrel{(i)}{=} \bigcap_{\omega\in \conv \Lambda } P(\omega) &\stackrel{(ii)}{=} \bigcap_{\omega\in \overline{\conv \Lambda }} P(\omega) \\
 & = \bigcap_{\omega\in \conv\pth{\ext\pth{\overline{\conv \Lambda}}}} P(\omega)\stackrel{(i)}{=} \bigcap_{\omega\in \ext\pth{\overline{\conv \Lambda }}} P(\omega).
 \end{aligned}\qedhere\]
\end{proof}

\noindent
This has an interesting computational consequence, which is already known as a tractable instance of static robust optimization~\cite[Chapter 1, Corollary~1.3.5~(i)]{ben2009robust}, and which we include here for completeness.

\begin{corollary}\label{cor:one-point-hit}
  Let $\Omega \subseteq \RR^p$ be a polytope with $v$ vertices and let
  $P(\Omega)$ be an affine family of polyhedra in $\RR^d$, with every
  member of the family defined by $m$ constraints. Deciding the existence of a one-point hitting set for $P(\Omega)$ reduces to solving
  a linear program with $d$ variables and $v\cdot m$ constraints.
\end{corollary}
\begin{proof}
We apply Lemma~\ref{lem:extreme}~(i) to the set $\Lambda=\ext(\Omega)$. Since $\conv (\ext(\Omega))=\Omega$, it comes that $\bigcap_{\omega\in \ext(\Omega)} P(\omega) = \bigcap_{\omega\in \conv (\ext(\Omega))} P(\omega)= \bigcap_{\omega\in \Omega} P(\omega) $.
That is to say, $P(\Omega)$ admits a one-point hitting set if and only if the set $P(\ext(\Omega))=\{P(\omega)\colon \omega \in \ext(\Omega)\}$ admits a one-point hitting set. The latter condition is equivalent to the existence of $\vc x\in\RR^d$ such that for all $\omega \in \ext(\Omega)$, $A(\omega)\vc x \le b(\omega)$. This is equivalent to finding a feasible solution of a linear program  with $d$ variables and $v\cdot m$ constraints.
\end{proof}

\subsection{Structure of a minimum-size hitting set.}

Let $\alpha,\beta \in \RR$ and let $P([\alpha,\beta])$ be an affine
family of polytopes. By Lemma~\ref{lem:extreme}, for any point $\vc{x}
\in \RR^d$, the set $\dual P(\vc x) = \{\omega \in \RR \colon x \in
P(\omega)\}$ is a closed interval. For $\lambda \in [\alpha,\beta]$
let us define 
\[ \sigma_P(\lambda) = \sup \ \{\lambda\}\cup\left\{\nu \in [\alpha,\beta] \colon \exists \vc{x} \in \RR^d
\text{ s. t. } [\lambda,\nu] \subseteq \dual P(\vc x)\right\}.\]
Note that if $P(\lambda)=\emptyset$, then $\sigma_P(\lambda)=\lambda$. We argue that this supremum is always attained.\footnote{Let us stress that if $P([\alpha,\beta])$ is an affine family of {\em polyhedra}, then the supremum $\sigma_P(\lambda)$ may be finite but not attained in the sense that $P([\lambda, \sigma_P(\lambda)])$ may not have a one-point hitting set. Indeed, consider for example $[\alpha,\beta] = [0,2]$ and $P(\omega)=\{x\in \RR \colon x(\omega-1) \ge 1 \}$; For $\omega\in [0,1)$, we have $P(\omega)=(-\infty,\frac{1}{\omega-1}]$, so that $\sigma_P(0)=1$; Note, however, that $P(1)=\emptyset$, which prevents $P([0,1])$ from having a one-point hitting set.}

\begin{lemma}\label{lem:cpct-hit}
  For every $\lambda \in [\alpha,\beta]$, either $P(\lambda)$ is empty
  or there exists $x_\lambda \in \RR^d$ such that $\dual P(x_\lambda)$
  contains $[\lambda, \sigma_P(\lambda)]$.
\end{lemma}
\begin{proof}
  Let $\lambda \in [\alpha,\beta]$. We can assume that $P(\lambda)$ is nonempty, as otherwise the statement is trivial. We can also assume that $\sigma_P(\lambda)>\lambda$, as otherwise any point $x_\lambda \in P(\lambda)$ satisfies the condition. Now, suppose, by contradiction, that $\cap_{t \in [\lambda,\sigma_P(\lambda)]} P(t) = \Hit{[\lambda,\sigma_P(\lambda)]}$ is empty. Since $\sigma_P(\lambda)>\lambda$, we have $\Hit{[\lambda,\sigma_P(\lambda))} = \Hit{[\lambda,\sigma_P(\lambda)]}$ by Lemma~\ref{lem:extreme}~(ii), so $\cap_{t \in [\lambda, \sigma_P(\lambda))} P(t)$ is already empty. By Helly's theorem, there exist $t_1 \le t_2 \le \ldots \le t_{d+1}$ in $[\lambda, \sigma_P(\lambda))$ such that $\cap_{i=1}^{d+1} P(t_i)$ is empty. (Note that Helly's theorem holds for infinite families of {\em compact} convex sets.) Yet, $t_{d+1} < \sigma_P(\lambda)$ ensures that there exists $x \in \RR^d$ such that $[\lambda, t_{d+1}] \subseteq \dual P(x)$. In  particular, $x \in \cap_{i=1}^{d+1} P(t_i)$, a contradiction.
\end{proof}

\noindent
Let us write $\sigma_P^{(i)} = \underbrace{\sigma_P \circ \sigma_P \circ \ldots \circ \sigma_P}_{i \text{ times}}$. Lemma~\ref{lem:cpct-hit} yields a simple characterization of the size of hitting sets of one-parameter families of polytopes.

\begin{lemma}\label{lem:greedy}
  An affine family of polytopes $P([\alpha,\beta])$ has a hitting set
  of size $k$ if and only if $\sigma_P^{(k)}(\alpha) = \beta$.
\end{lemma}
\begin{proof}
  Suppose that $\sigma_P^{(k)}(\alpha) = \beta$.
  Let $t_0 = \alpha$ and for $i=1, 2, \ldots k$ let $t_i = \sigma_P(t_{i-1})$.
  By Lemma~\ref{lem:cpct-hit}, for $i=1, 2, \ldots, k$ there exists $x_i \in \RR^d$ such that $\dual P(x_i) \supseteq [t_{i-1},t_i]$.
  Since $t_k = \sigma_P^{(k)}(\alpha) = \beta$, we have $\cup_{i=1}^k [t_{i-1},t_i] = [\alpha,\beta]$, and $\{x_1,x_2, \ldots, x_k\}$ is a hitting set for $P([\alpha,\beta])$.

  Conversely, suppose that $P([\alpha,\beta])$ admits a hitting set $H = \{x_1,x_2, \ldots, x_k\}$.
  Let us write $\dual P(x_i) = [\ell_i,r_i]$ and assume, without loss of generality, that $\ell_1 \le \ell_2 \le \ldots \le \ell_k$.
  Observe that, by definition, $r_i=\min\{\beta, \sigma_P(\ell_i)\}$, hence $r_i\le \sigma_P(\ell_i)$ for all $i\in [k]$.
  Also observe that $\ell_i\le r_{i-1}$, since otherwise $H$ misses $P(\omega)$ for some $\omega\in (r_{i-1},\ell_i)$.
  We must also have $\beta\le r_k$, since otherwise $H$ misses $P(\beta)$, and $\ell_1\le\alpha$, otherwise $H$ misses $P(\alpha)$.
  Combining these observations, and using the fact that the function $\sigma_P$ is nondecreasing, we obtain:
\[
 \beta \le r_k \le \sigma_P(\ell_k) \le \sigma_P(r_{k-1}) \le \sigma_P(\sigma_P(\ell_{k-1}) \le\ldots \le \sigma_P^{(k)}(\ell_1) \le \sigma_P^{(k)}(\alpha).
\]
  Since $\sigma_P(\lambda)$ is at most $\beta$, we must have $\sigma_P^{(k)}(\alpha) = \beta$, as claimed.
\end{proof}

\subsection{Parametric search on a linear programming algorithm.}

Lemma~\ref{lem:greedy} reduces the computation of a minimum-size
hiting set for a one-parameter affine family of polytopes $P([\alpha,\beta])$ to the
computation of the function $\sigma_P$. We now focus on the latter
problem.

\subsubsection{Comparing is easy}

Let us fix some real $s$ and write $s^* = \sigma_P(s)$; we assume that
$s$ is known, and want to determine $s^*$. We first notice that we
already have efficient algorithms for comparing any input real $t$ to
$s^*$: this is equivalent to deciding whether $P([s,t])$ can be hit by
a single point, and therefore reduces to testing whether $P(s) \cap
P(t)$ is empty, by Corollary~\ref{cor:one-point-hit}, which can be
done by solving a linear program with $d$ variables and $2m$
constraints. Let us note that several deterministic algorithms have
been proposed to solve such linear program in $O(m)$ time when the
dimension $d$ is fixed~\cite{megiddo1984linear,Clarkson86,Dyer86,agarwalST1993,ChazelleM96,BronnimannCM99,Chan18}.

\subsubsection{From comparing to computing}

It turns out that a general technique, called {\em parametric search}~\cite{megiddo1983applying},
can turn\footnote{A more effective technique for this purpose was proposed by Chan~\cite{DBLP:journals/dcg/Chan99}, but unfortunately we could not apply it here.} an algorithm for {\em comparing to} $s^*$ into an algorithm
for {\em computing} $s^*$, the latter having complexity $O(f(n)^2)$
where $f(n)$ is the complexity of the former. This suggests that all
we need is to apply parametric search to one of these linear-time
algorithms for linear programming to get a quadratic-time algorithm to
compute $\sigma_P$. There is, however, a catch: parametric search can
only be applied to a comparison algorithm that can be modeled by an
algebraic decision tree where the decisions have bounded degree. This
is, fortunately, the case of Megiddo's
algorithm~\cite{megiddo1984linear} (a fact more easily checked from
Clarkson's summary of that algorithm~\cite{Clarkson86}). For completeness, we
recall in Appendix~\ref{app:adt} the model of algebraic decision
trees and summarize in Appendix~\ref{app:ps} the parametric search
technique.

\subsection{Proof of Theorem~\ref{thm:onepara}}

Let us first summarize how our algorithm works in the real RAM model.

\begin{proposition}\label{prop:oneparaReal}
  For every constant $d$, there is a Real-RAM algorithm that computes
  in time $O(km^2)$, given an affine family of polytopes
  $P([\alpha,\beta])$, each defined by at most $m$ constraints, the
  size $k$ of the smallest hitting set of $P([\alpha,\beta])$.
\end{proposition}
\begin{proof}
  We are given as input $P([\alpha,\beta])$. We put $\alpha_0 =
  \alpha$ and compute $\alpha_i = \sigma_P(\alpha_{i-1})$, for $i=1,2,
  \ldots$, until we reach some value $\alpha_i \ge \beta$. At this
  point, we return $i$ as the size of the smallest hitting set of
  $P([\alpha,\beta])$. We compute $\alpha_i$ from $\alpha_{i-1}$ by
  performing parametric search on the algorithm $\A$ that, given some
  real $t$, uses Megiddo's algorithm to solve the linear program that
  determines whether $t \le \sigma_P(\alpha_{i-1})$. Since the
  dimension $d$ is fixed, Megiddo's algorithm takes $O(m)$ time to
  solve one comparison to $\sigma_P(\alpha_{i-1})$, and the
  computation of one $\alpha_i$ takes $O(m^2)$ time. Altogether, the
  algorithm takes $O(km^2)$ time, where $k$ is the size of the
  smallest hitting set of $P([\alpha,\beta])$.
\end{proof}

To prove Theorem~\ref{thm:onepara}, it remains to analyze the bit complexity of the numbers manipulated by the algorithm of Proposition~\ref{prop:oneparaReal} when $k$ is fixed in addition to the ambient dimension $d$ of the polytopes. This proof relies on the observation that if $\A$ is an algebraic decision tree of height $h$ that decides a monotone decision problem $P(\cdot)$, then the algorithm obtained by performing parametric search on $\A$ can also be modeled by a tree of height $O(h^2)$, which is almost an algebraic decision tree (except for the leaves, that do not output only true or false). The key ingredient for this is the following lemma.

\begin{lemma}\label{app:rootadt}
  For every constant $d_1$ and $d_2$, there exist algebraic decision trees of constant complexity that solve the following problems given an integer $r \le d_1$ and a vector $V \in \RR^{d_1+d_2+2}$ describing the coefficients of two univariate polynomials $p_1(X)$ and $p_2(X)$ of degree at most $d_1$ and $d_2$, respectively: 
  \begin{itemize}
  \item[(i)] does $p_1(X)$ have at least $r$ real roots? 
  \item[(ii)] is the value of $p_2$ in the $r$-th root of $p_1$ strictly positive (resp. strictly negative, zero)?
  \end{itemize}
 \end{lemma}
\begin{proof}
  Questions~(i) and~(ii) can be formulated as boolean formulas in a constant number of variables. Let $\pi_s(V,X)$ denote $p_s(X)$ for $s=1,2$. For (i),  the polynomial $p_1$ has at least $r$ distinct real roots $(\alpha_1<\cdots<\alpha_r)$ if and only if the following formula is satisfiable
  \[
  \exists (x_1,x_2,\ldots,x_{d_1})\in \RR^{d_1},\left(\bigwedge_{\ell=1}^{d_1} \pth{\pi_1(V, x_\ell)=0 \vee (\ell > r)}\right)  \wedge \left(\bigwedge_{\ell=1}^{d_1-1}(x_{\ell+1}>x_\ell)\right).
  \]
  For (ii), we treat the case of strictly positive sign, the two other cases follow the same path. The polynomial $p_1$ has at least $r$ distinct real roots $(\alpha_1<\cdots<\alpha_r)$, all the roots of $p_1$ are contained in $[\alpha_1,+\infty)$ and the evaluation of $p_2$ in the $r$-th root $\alpha_r$ of $p_1$ is strictly positive if and only if the following formula is satisfiable
\[\begin{aligned}
\exists & (x_1,x_2\ldots, x_{d_1})\in\RR^{d_1},\forall x\in\RR,\\
& \quad \pth{\bigwedge_{\ell=1}^{d_1} \pth{\pi_1 (V,x_\ell)=0 \vee (\ell > r)}\right)  \wedge \left(\bigwedge_{\ell=1}^{d_1-1}(x_{\ell+1}>x_\ell)} \\
&  \wedge \pth{(\pi_1(V,x)=0) \Rightarrow \pth{\bigvee_{\ell=1}^{d_1} \pth{(x=x_\ell) \wedge (\ell \le r)}}\vee \pth{\bigwedge_{\ell=1}^{d_1}\pth{(x>x_\ell) \vee (\ell > r)}}}\\
 &\wedge  \bigwedge_{\ell=1}^{d_1}\pth{\pi_2(V,x_\ell)>0 \vee (\ell \neq r)}.
\end{aligned}
\]
By Theorem~\ref{thm:elim}, these formulas are equivalent to quantifier free formulas in the input vector $V$ and $r$, involving a constant number of polynomials of constant degree. These formulas can in turn be modeled by algebraic decision trees of constant complexity. 
\end{proof}

We can now complete the proof of Theorem~\ref{thm:onepara}.

\begin{proof}[Proof of Theorem~\ref{thm:onepara}]
  Let $V$ be a vector of real numbers consisting of the parameters that represent $P([\alpha,\beta])$. Let $\A_0$ denote an algebraic decision tree that models the comparison, using Megiddo's linear-time algorithm, of $t$ to $\sigma_P(s)$. The input of $\A_0$ is $s$, $V$ and $t$. The height of $\A_0$ is $O(m)$. Here is a key observation: for any constant $d_1$, $d_2$, and $i$ there exists a constant-size algebraic decision tree that takes as input the vector of coefficients of two univariate polynomials $\Pi_1$ of degree $d_1$ and $\Pi_2$ of degree $d_2$ and decides the sign of $\Pi_1$ on the $i$th real root of $\Pi_2$. (See Lemma~\ref{app:rootadt} in Appendix~\ref{app:ps}).
  
  Now, consider the real RAM algorithm that computes $\sigma_P(s)$ by performing parametric search on $\A_0$. This algorithm has complexity $O(m^2)$ and takes $s$, $V$ and $t$ as input. Observe that the value that it outputs is not an arbitrary real: the principles of parametric search ensure that the output is either $+\infty$ or a real root of a polynomial $\Pi_1$ determining the branching in one of the nodes of $\A_0$ (see Appendix~\ref{app:ps}). We can therefore represent the output of the parametric search by an integer (the number of the root) and a polynomial of constant degree and whose coefficients are constant degree polynomials of some input coordinates. We can model this parametric search by a tree $\A_1$ that is almost an algebraic decision tree: $\A_1$ takes $s$ and $V$ as input, and each leaf contains the description of $\sigma_P(s)$ in the form of the integer $i$ and the polynomial $\Pi_1$. The only aspect in which $\A_1$ is not an algebraic decision tree is that it does not solve a decision problem.

  We can now append at each leaf of $\A_1$ a second copy of $\A_1$ where the input parameter $s$ has been substituted for the $i$th root of $\Pi_1$. This only requires changing, in the second copy of $\A_1$, for branching polynomial $\Pi$, the one-node evaluation of the sign of $\Pi$ in $s$ by a constant-size subtree evaluating the sign of $\Pi$ in the $i$th root of $\Pi_1$. This result in a tree $\A_2$ that takes $s$ and $V$ as input, and where each leaf contains the description of $\sigma_P^{(2)}(s)$. Again, the principles of parametric search (see Appendix~\ref{app:ps}) ensures that the output can be presented in the form of the integer $i$ and a polynomial $\Pi_2$ of constant degree, the coefficients of which are constant degree polynomials of some input coordinates. Again, the aspect in which $\A_2$ is not an algebraic decision tree is that it does not solve a decision problem. Again, $\A_ 2$ has height $O(m^2)$.

  And so on, for every constant $i$ we can construct a tree $\A_i$ of height $O(m^2)$, that takes $s$ and $V$ as input, and where each leaf contains the description of $\sigma_P^{(i)}(s)$. Finally, in $\A_k$, we substitute every leaf by a constant-size algebraic decision tree that compares the number stored in that leaf to the input number $\beta$. The resulting tree $\A$ is an algebraic decision tree of height $O(m^2)$ that takes $s$ and $V$ as input, and that compares $\sigma_P^{(k)}(s)$ to $\beta$. Executing this tree for $s=\alpha$ gives a strongly quadratic algorithm for deciding whether $P([\alpha,\beta])$ has a $k$-point hitting set.
\end{proof}

Proving Corollary~\ref{cor:appli} now amounts to going from a decision problem to the associated optimization problem.

\begin{proof}[Proof of Corollary~\ref{cor:appli}]
  With Lemma~\ref{lem:decision-hit-nofd} and Theorem~\ref{thm:onepara}, we already have that for every constant $d$ and $k$, there is an algorithm that solves the decision version of the $k$-adaptability Problem~\eqref{eq:fa} in the special case where $\ell=0$ and $\Omega$ is an interval, in time strongly quadratic in $m$. That algorithm can be modeled by an algebraic decision tree (this is the final tree $\A$ in the proof of Theorem~\ref{thm:onepara}). We can therefore perform parametric search on that algorithm to solve the optimization version of these problems in time strongly quartic in $m$.  
\end{proof}

\section{Concluding remarks}

\begin{enumerate}
\item The quantifier elimination algorithms do not need the maps $A_f$, $A_s$, $b$, and $c_s$ (for the finite adaptability) or the maps $A$ and $b$ (for families of polyhedra) to be affine, but merely to be polynomial of constant degree. Theorems~\ref{thm:appli} thus generalizes, {\em mutatis mutandis}, to this setting. Note that the assumption that these maps are affine is used in the correspondence between the finite adaptability and the hitting/covering of affine families of polyhedra, in the hitting/covering duality, and in the proof of Theorem~\ref{thm:onepara} and Corollary~\ref{cor:appli}.

\item The assumption in Theorem~\ref{thm:onepara} that $P$ be an affine family of polytopes, rather than polyhedra, is used to ensure that $\sigma_P$ is not only a supremum, but actually a maximum (this underlies the proof of Lemma~\ref{lem:cpct-hit}). We conjecture that if $P$ is a one-parameter family of polyhedra and $P([\lambda, \sigma_P(\lambda)]$ cannot be hit by a single point, then the polyhedron $P(\sigma_P(\lambda))$ must be empty. If that is true, then Theorem~\ref{thm:onepara} readily generalizes to affine families of polyhedra.
\item A natural question is whether the hitting set problem for affine families of polyhedra is fixed-parameter tractable with respect to the dimensions (of the ambient space of the polyhedra as well as the parameter space).
\end{enumerate}

\section*{Aknowledgements}The authors thank Safia Kedad-Sidhoum, Anton Medvedev and Frédéric Meunier for helpful discussions on finite adaptability and Guillaume Moroz for helpful discussions on computer algebra.
\bibliographystyle{plain}
\bibliography{bibliography}


\appendix

\section{Strongly polynomial algorithms}\label{app:spa}

The standard model of computation in computational geometry is the \emph{Real-RAM} model, a variant of the classical random access machine (RAM) that operates on arbitrary real numbers instead of finite binary words.
A succinct formal presentation of that model can be found in Erickson~\cite[$\mathsection 7.4$]{ericksontree}, and Erickson et al.~\cite[$\mathsection 6.1$]{smoothing}.
The computational power of this model is considerable, 
(see the discussion in~\cite[$\mathsection 1$]{smoothing}).
To ensure that running a polynomial-time Real-RAM algorithm on a finite-description input yields a polynomial-time algorithm in the usual (\textit{e.g.} Turing machine) sense,
one must check that the cost of each elementary operation remains polynomial, hence that the bitsize of the numbers manipulated does not blow up.

Algorithms that run in polynomial time on a Real-RAM \textit{and} that translate to polynomial-time algorithms in the usual sense are called \emph{strongly polynomial}.
Whether linear programming can be solved in strongly polynomial time remains a major open problem, known as Smale's 9th problem~\cite{Smale98}.

\section{Algebraic decision trees}\label{app:adt}

The \emph{algebraic decision tree} model of computation~\cite{SY82} represents an algorithm with input $x \in \RR^n$ as a rooted tree where (i) each internal node $v$ is decorated with a polynomial $p_v$ with constant coefficients and with variables some of the input $x_i$, (ii) for each internal node, the edges towards its descendants are labeled by sign conditions ($<0$, $\le 0$, $=0$, $\neq 0$, $\ge 0$ or $>0$) that are  mutually exclusive and cover all cases, and (iii) each leaf has a label that serves as answer to the algorithm. To execute an algebraic decision tree $T$ on an input $x\in \RR^n$, we traverse $T$ starting from the root, determining at each node $v$ the sign of the polynomial $p_v$ on the input $x$ and taking the corresponding edge towards a descendant, and return the label of the leaf we eventually reach. Let us stress that this model of computation is nonuniform, in the sense that the tree is actually a family of trees, one per input size~$n$. Let us also emphasize that any polynomial Real RAM algorithm that can be modeled by an algebraic decision tree is strongly polynomial, as the arithmetic operations only evaluate polynomials of constant degree in some input parameters.

\section{Parametric search}\label{app:ps}

Let $P(\lambda)$ be a decision problem depending on a real parameter $\lambda$.
Suppose that $P$ is \emph{monotone} in the sense that if $P(\lambda)$ is true, then $P(\mu)$ is also true for all $\mu \le \lambda$.
Then, there exists some $\lambda^* = \sup \{\lambda \in \RR \colon P(\lambda) \text{ is true}\}$, possibly $\infty$.
The parametric search technique ~\cite{megiddo1983applying} (see also~\cite{agarwal1998efficient}) is a method that can turn an algorithm  $\A$ that evaluates $P(\lambda)$ given $\lambda$ into an algorithm that computes $\lambda^*$.
This technique requires that $\A$ be presented as an algebraic decision tree.
Intuitively, it consists in identifying the path in $\A$ that would be followed if that tree was executed on $\lambda^*$.

Let us sketch a simple version of that technique that is sufficient for our purpose.
We traverse the tree while maintaining an interval $I$ known to contain $\lambda^*$, initially set to $\RR$.
When reaching a node $v$, we check whether the sign of the polynomial $p_v$ is constant over $I$.
If it is, we follow the suitable branch and keep $I$ unchanged.
Otherwise, we compute the set $S$ of roots of $p_v$ and use $\A$ to determine $P(\lambda)$ for each $\lambda \in S$; the monotony ensures that the largest root $r$ of $p_v$ such that $P(r)$ is true and the smallest root $s$ of $p_v$ such that $P(s)$ is false are consecutive in $S$; they determine our new interval $I$, and we follow the branch indicated by the sign of $p_v$ on $I$.
We continue this procedure until we reach a leaf, and return the upper endpoint of $I$.
Altogether, if $h$ is the height of the tree, the parametric search takes $O(h)$ calls to $\A$ on specific values as well as $O(h)$ additional computation.
In other words, if $\A$ has complexity $O(f(n))$, then the parametric search determines the exact value of $\lambda^*$ in time $O(f(n)^2)$.

\section{Comparison between two-stage robust optimization and $k$-adaptability}\label{app:twostage}

The "General Decision" Theorem deals with deciding the truth or falsity of a first order formula with no free variables. It will be useful in our context since the two-stage robust optimization problem as well as the $k$-adaptability can be formulated within this frame.
\begin{theorem}{\cite[$\mathsection 14$, Theorem~14.14]{basu:hal-01083587}}\label{thm:gendec}
  Let $\phi$ be a first-order formula with no free variables of the form $(Q_1 \mathbf{X}_1)(Q_2 \mathbf{X}_2)\ldots(Q_n \mathbf{X}_n)F(X)$, where $Q_1$, $Q_2$, \ldots, $Q_n$ are quantifiers in $\{\forall,\exists\}$ and $F$ is a quantifier free formula. Assume that each set $\mathbf{X}_{i}$ has $b_i$ variables and that $F$ is built with at most $s$ polynomials of degree at most $r$. 

  \begin{itemize}
  \item[(i)] The truth of the sentence $\phi$ can be decided in $\nu$ arithmetic operations in the ring generated by the coefficients of
    the polynomials, and
  \item[(ii)] if the coefficients of the input polynomials are in $\mathbb{Z}$ and have bitsize at most $\tau$, then the polynomials that arise in the algorithm have coefficients of bitsize bounded by $\tau \delta$,
  \end{itemize}
  where $\nu = s^{(b_1+1)(b_2+1)\cdots(b_n+1)}r^{O(b_1)O(b_2)\cdots O(b_n)}$ and $\delta = r^{O(b_1)O(b_2)\cdots O(b_n)}$.
\end{theorem}

We now apply this theorem to the decision versions of the two-stage problem and the $k$-adaptability problem.
Let $P_t(\Omega)$ be the affine family of polyhedra defined by
  \[
  P_t(\omega) = \left\{
  (x_f, x_s) \in \RR^{\df+\ds} \colon
 	\pth{\begin{matrix} 
 		A_f(\omega) & A_s(\omega)\\ 
 		{c_f}^T & c_s(\omega)^T
 	\end{matrix}} 
 	\pth{\begin{matrix}
 		\vc{x_f}\\
 		\vc{x_s} 
 	\end{matrix}} 
 		\le 
 	\pth{\begin{matrix} 
 		\vc b(\omega)\\
 		t
 	\end{matrix}}
  \right\}.
  \]
\subsection{General Decision for the two-stage problem}
The two-stage robust optimization Problem~\ref{eq:ts} in its decision version is equivalent to deciding the truth of the following sentence
\[\Phi \equiv  \exists x_f \in \RR^\df, \forall \omega \in \RR^p, \exists x_s \in \RR^\ds \quad \phi(x_f,\omega,x_s)
	\]
where
  \[
    \phi(x_f,\omega, x_s) \equiv 
    \omega\not\in\Omega \vee 
    (x_f,x_s)\in P_t(\omega).
  \]
We apply Theorem~\ref{thm:gendec} with parameters $s\le v+m+1$, $r\le 2$, $n=3$, $b_1=\df$, $b_2=p$, $b_3=\ds$. This leads to an arithmetic complexity of $(v+m+1)^{(\df+1)(p+1)(\ds+1)}2^{O(\df)O(p)O(\ds)}$ that is \emph{$O\pth{(v+m)^{(\df+1)(p+1)(\ds+1))}}$} to solve the decision problem of the two-stage optimization problem when $\df,\ds,p$ are fixed.
\subsection{General Decision for the $k$-adaptability problem}
The $k$-adpatability Problem~\ref{eq:fa} in its decision version is equivalent to deciding the truth of the following sentence
\[
  \Psi \equiv \exists x_f \in \RR^{\df}, (x_s^1, x_s^2,\ldots ,x_s^k)\in (\RR^\ds)^k, \forall\omega\in\RR^p\quad \psi(x_f,x_s^1,x_s^2,\ldots,x_s^k,\omega) 	
  \]  
  where
  \[
    \psi(x_f,x_s^1,x_s^2,\ldots,x_s^k,\omega) 	 \equiv \omega\not\in\Omega \vee 
    \bigvee_{i\in [k]} \pth{(x_f,x_s^i)\in P_t(\omega)}.
  \]
We apply Theorem~\ref{thm:gendec} with parameters $s\le v+k(m+1)$, $r\le 2$, $n=2$, $b_1=\df+k \ds$, $b_2=p$. This leads to an arithmetic complexity of $(v+k(m+1))^{(\df+k\ds +1)(p+1)}2^{O(\df +k\ds)O(p)}$ that is \emph{$O\pth{(v+m)^{(\df+k\ds +1)(p+1)}}$} to solve the decision problem of the $k$-adaptability when $k,\df,\ds,p$ are fixed. 

We see from this comparison that solving the $k$-adaptability is faster than solving the two-stage robust optimization problem as soon as $k$ is fixed and verifies $k\le \df  +1$.
\end{document}